# Quantitative MRI
# – Absolute $T_1$, $T_2$ and Proton Density Parameters from Deep Learning

Qing Lyu, Ge Wang
Biomedical Imaging Center
CBIS/BME/SoE
Rensselaer Polytechnic Institute
Troy, NY, USA

*Abstract* – Quantitative MRI is highly desirable in terms of intrinsic tissue parameters such as $T_1$, $T_2$ and proton density. This approach promises to minimize diagnostic variability and differentiate normal and pathological tissues by comparing tissue parameters to the normal ranges. Also, absolute quantification can help segment MRI tissue images with better accuracy compared to traditional qualitative segmentation methods. Currently, there are several methods proposed to quantify tissue parameters; however, all of them require excessive scan time and thus are difficult to be applied in clinical applications. In this paper, we propose a novel machine learning approach for MRI quantification, which can dramatically decrease the scan time and improve image quality.

## I. Introduction

Magnetic resonance (MR) techniques including magnetic resonance spectroscopy (MRS) and magnetic resonance imaging (MRI) have been widely used world-wide after over 40 years' major innovation and rapid development. However, in practice most often used MR techniques are restricted in qualitative or weighted interpretation and lack of quantitative analysis. In modern clinical MRI, when comparing a tissue or material with their surroundings, it is typically referred to as being 'hyperintense' or 'hypointense' without thorough quantitative analysis. As a result, MRI may fail to provide accurate indication of the severity of diseases. In recent years, several methods for quantitative analysis of MRI signals were developed, such as the measurement of the longitudinal relaxation time ($T_1$) [1-5] and the transverse relaxation time ($T_2$ and $T_2^*$) [3, 5-7]. Methods for measuring the proton density ($\rho$) were also proposed [5]. However, most of these quantitative methods can only provide information on a single parameter at a time, and they require long scan time to acquire MR signals for a good signal-to-noise ratio, which compromises their clinical utility.

An approach called MR fingerprinting (MRF) was proposed in 2013, which is a technique that could theoretically be applied to extract intrinsic tissue parameters quantitatively [8]. MRF uses a randomized acquisition strategy that results in signals from different tissues with unique signal evolution trajectories or 'fingerprints'. The signal post-processing involves a pattern recognition algorithm that can match those fingerprints to a predefined dictionary, translating the fingerprints into quantitative maps of the magnetic parameters of interest.

In this paper, we propose a new approach for quantitative analysis MRI signals in the machine learning framework. According to our scheme, we achieve a quantitative mapping of intrinsic tissue parameters (that is, $T_1$, $T_2$ and $\rho$) directly from acquired MRI signals, greatly decreasing the signal acquisition time and optimizing image quality in the data-driven manner.

## II. Methodology

Our deep learning approach targets a data-driven quantitative mapping from sampled MRI data to an intrinsic tissue parameter matrix. The overall idea is illustrated in Figure 1. This workflow includes two main parts: (1) the MRI signal data generation part for generating $T_1$, $T_2$ and $\rho$-weighted signals and (2) the neural network part for mapping MRI signals to intrinsic tissue parameters.

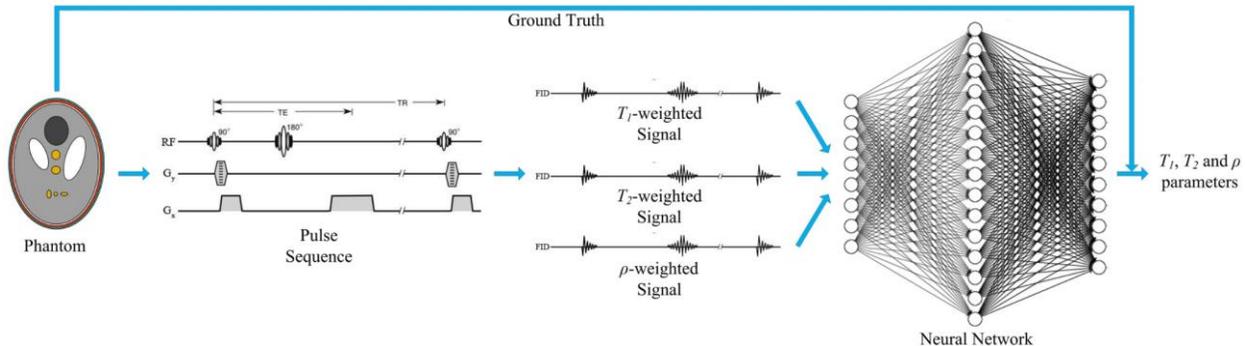

Figure 1. Quantitative MRI approach in the machine learning framework.

### A. Data Generation

Generalized Shepp-Logan phantoms were used to demonstrate the feasibility and merits of our proposed approach. Totally, 51,000 2D phantom images were generated for this pilot study. For details please see [9]. 50,000 phantoms were randomly selected for training, and the remaining 1,000 phantoms were used for testing.

MRI signals were produced with the popular spin-echo pulse sequence. The Bloch equation governs this data generation process. The detailed description of this process can be seen in [9]. Gaussian noise was added into the simulated MRI signals. For generating $T_1$-weighted MRI signals, the repetition time (TR) was set to 500 ms and the echo time (TE) to 15 ms. For $T_2$-weighted MRI signals, the TR and TE were set to 10,000 ms and 300 ms respectively. For $\rho$-weighted MRI signals, TR and TE were 10,000 ms and 15 ms respectively. For each phantom, $T_1$, $T_2$, and $\rho$-weighted signals were alternatively generated: the first spin-echo frequency-encoding period generated $T_1$-weighted signals, the second spin-echo frequency-encoding period generated $T_2$-weighted signals, and the third spin-echo frequency-encoding period generated $\rho$-weighted signals, then next three consecutive periods repeated this order, so on and so forth. The sampling frequency was 10,000 Hz, sampling 64 data points during each spin-echo frequency-encoding period. After sampling, mixed $T_1$, $T_2$, and $\rho$-weighted signal data were put into a 64 × 64 matrix as the input to a reconstruction network.

### B. Neural Network

The architecture of our neural network for MRI image reconstruction is shown in Figure 2. It consists of 6 convolutional layers and two fully-connected layers. The input of the neural network was the sampled signal matrix with the size of 64 × 64 complex-valued numbers. Then, the signal matrix was expanded to two 64 × 64 channels through splitting each complex number into two real numbers representing real and imagery parts respectively. The first two convolutional layers each had 64 filters with kernel size of 5 × 5. The third convolutional layer had only one filter with 5 × 5 kernel size. After the third convolutional layer, the outputs were vectorized with the length of $64^2$. After the second fully-connected layer, the output vectors were reshaped into 64 × 64 × 1 tensors and then there were the fourth and fifth convolutional layers with 64 filters of 5 × 5 kernel size. Finally, the sixth convolutional layer had three filters, giving the final output as a 64 × 64 × 3 matrix. The ground truth was a 64 × 64 × 3 matrix that 3 channels containing the ground truth $T_1$, $T_2$ and $\rho$ tissue parameters. The loss function was a combination of the mean-squared-error and the $L_1$-norm penalty, with a relaxation factor of 0.0001. This $L_1$-norm penalty was applied to the feature map of the final hidden layer. The RMSProp algorithm [10] was used with mini-batches of size 100, momentum 0.0 and decay 0.9. Training process continued for 500 epochs with the learning rate 0.0001 for the first 100 epochs and being then divided by 1.01 every two epochs. All the neural parameters were initialized in the way of [11]. The neural network was implemented in Tensorflow [12] on a NVIDIA GTX 1080Ti.

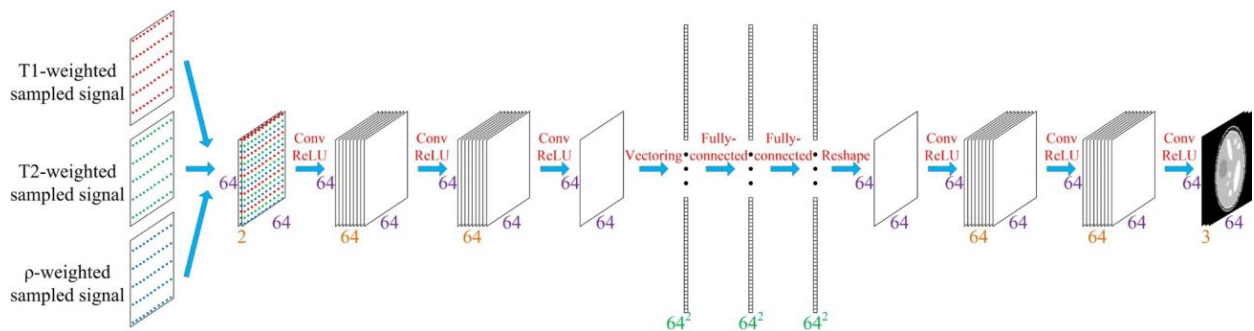

Figure 2. Architecture of the neural network for our proposed quantitative MRI.

### III. Numerical Results

In the numerical simulation, our neural network worked very well. Figure 3 shows the curves of MSE in the loss function during the training process. As the training process went on, the loss with either the training or testing dataset became gradually decreased. In Figure 4, the recovered tissue parameters were very close to their corresponding ground truth values, especially for the cerebrospinal fluid parameters. Quantitatively, as shown in Figure 5, the normalized MSE values (we normalized the maximum values for $T_1$, $T_2$ and $\rho$ to 1) decreased dramatically through the training process, $T_1$, $T_2$ and $\rho$-NMSE values for the 1,000 test phantoms were 0.00138, 0.00162 and 0.00125 after the first 50 epochs and declined to 0.00063, 0.00094 and 0.00058 after 500 epochs.

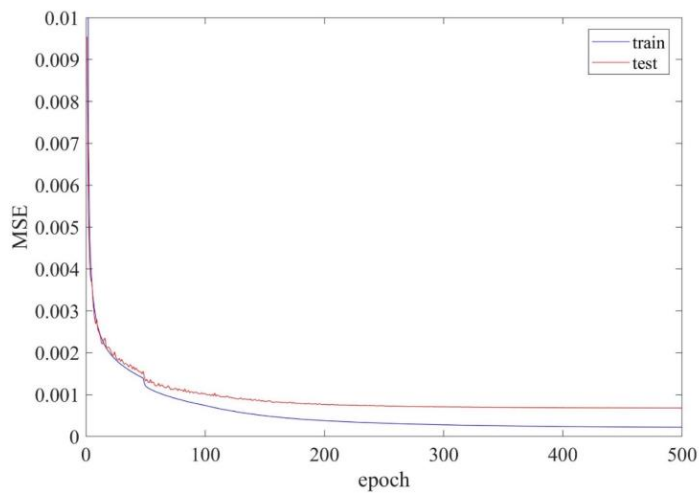

Figure 3. MSE loss reduction during the training process for training and testing data respectively.

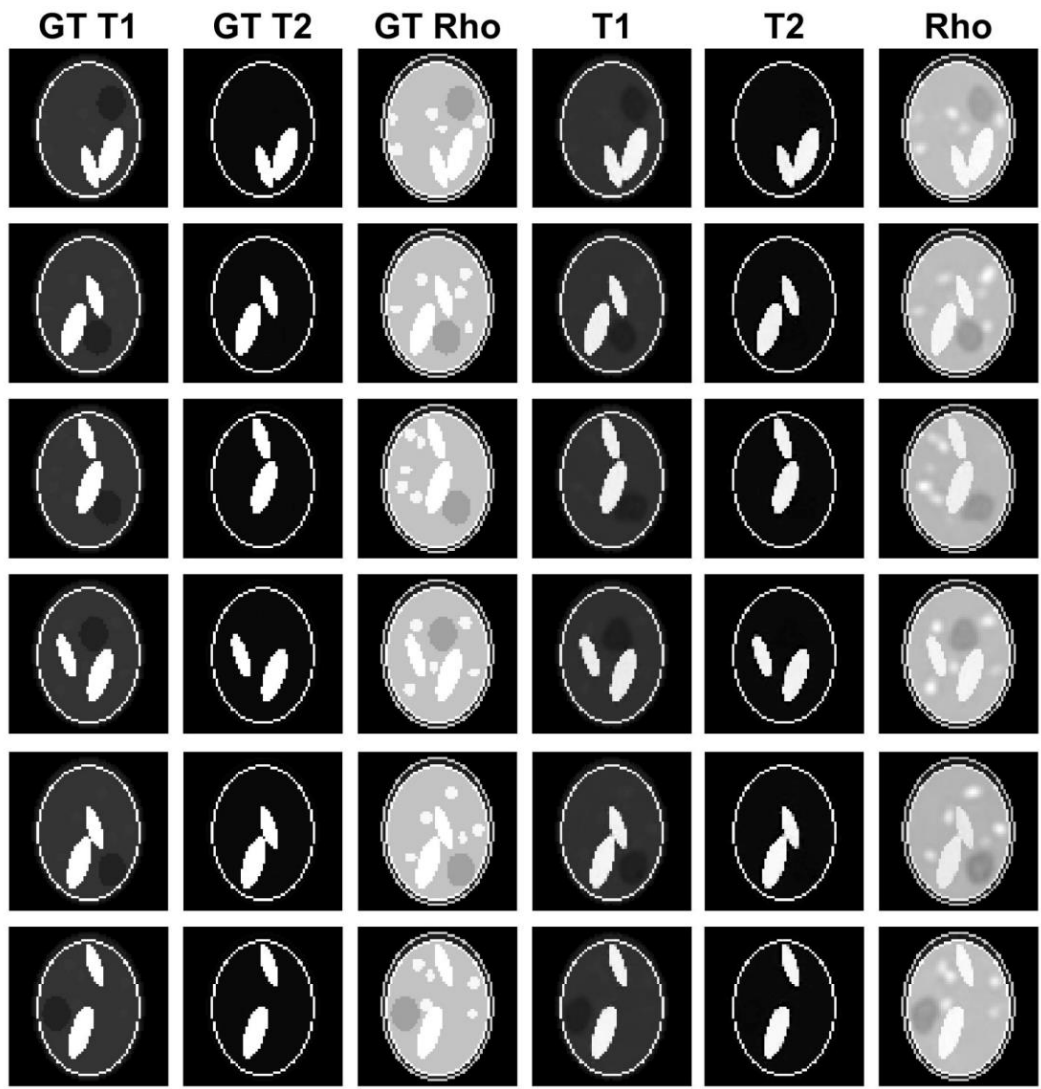

Figure 4. Recovered intrinsic tissue parameters after training over 500 epochs. Leftmost three columns: ground truth $T_1$, $T_2$, and $\rho$ parameters. Rightmost three columns: the corresponding recovered $T_1$, $T_2$ and $\rho$ parameters. Each row represents a phantom randomly selected.

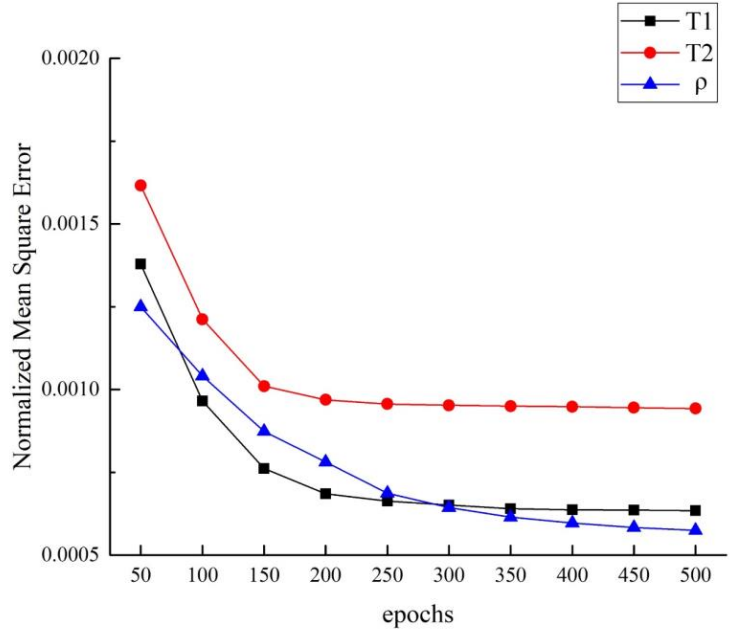

Figure 5. Mean squared errors for recovered $T_1$, $T_2$ and $\rho$ parameters for the 1,000 test phantoms.

## IV. Discussions and Conclusion

Our proposed data-driven method for quantitative MRI can directly recover intrinsic tissue parameters close to the ground truth tissue parameters from acquired MRI signals, which enables fast quantitative MRI/MRS analysis. As we alternatively generated $T_1$, $T_2$ and $\rho$-weighted MRI singles in 64 spin-echo frequency-encoding periods, the total signal acquisition time is not longer than a typical spin-echo acquisition method, suggesting that this approach is clinical applicable.